\documentclass[pre, preprint,groupedaddress, showkeys,showpacs]{revtex4}

\usepackage{graphicx}
\usepackage{fancyhdr}
\begin{document}


\title{Competition between ac driving-forces and L\'{e}vy flights in a nonthermal ratchet}


\author{Bao-quan  Ai$^{1}$}\email[Email: ]{aibq@hotmail.com}
 \author{Ya-feng He$^{2}$}

\affiliation{$^{1}$Laboratory of Quantum Information Technology,
ICMP and
 SPTE, South China Normal University, 510006 Guangzhou, China.\\
 $^{2}$College of Physics Science and
Technology, Hebei University, 071002 Baoding, China}


\date{\today}
\begin{abstract}
\indent Transport of overdamped particles in an asymmetrically
periodic potential is investigated in the presence of L\'{e}vy noise
and ac-driving forces. The group velocity is used to measure the
transport driven by the nonthermal noise. It is found that the
L\'{e}vy flights and ac-driving forces are the two different driving
factors that can break thermodynamical equilibrium. The competition
between the two factors will induce some peculiar phenomena. For
example, multiple transport reversals occur on changing the noise
intensity. Additionally, we also find that the group velocity as a
function of L\'{e}vy index is nonmonotonic for small values of the
noise intensity.
\end{abstract}

\pacs{ 05. 40. Fb, 05. 10. Gg, 05. 40. -a}



\maketitle
\section {Introduction}
\indent In systems possessing spatial or dynamical symmetry breaking,
Brownian motion combined with unbiased external input signals,
deterministic or random alike, can assist directed motion of
particles at submicron scales\cite{a1}. This subject was motivated
by the challenge to explain unidirectional transport in biological
systems\cite{a2}, as well as their potential technological
applications ranging from classical non-equilibrium models\cite{a3}
to quantum systems\cite{a4}. Ratchets have been proposed to model
the unidirectional motion driven by zero-mean non-equilibrium
fluctuations. Broadly speaking, ratchet devices fall into three
categories depending on how the applied perturbation couples to the
substrate asymmetry:  rocking ratchets\cite{a5}, flashing
ratchets\cite{a6}, and correlation ratchets \cite{a7}. Additionally,
entropic ratchets, in which Brownian particles move in a confined
structure, instead of a potential, were also extensively studied
\cite{a8}.  These ratchets demand three key ingredients \cite{a9}
which are (a)nonlinearity: it is necessary since the system will
produce a zero mean output from zero-mean input in a linear system;
(b)asymmetry (spatial and/or temporal): it can violate the symmetry of the response;
(c) fluctuating input zero-mean
force: it should break thermodynamical equilibrium.\\
\indent Most studies on ratchets have referred to the consideration
of normal diffusion driven by Gaussian noises. However, in the past
few years, anomalous diffusion has attracted growing attention,
being observed in various field of physics and related sciences
\cite{a10,a11,a12,a13}.
 Because the L\'{e}vy flights do not possess a finite mean-square displacement, their physical
significance has been ignored for a long time. However, in the
recent years, the growing experimental evidences suggest that there
is a need to consider a more general type of noise than Gaussian, i. e.,
L\'{e}vy noise. Description of physical models in terms of
L\'{e}vy flights becomes more and more popular
\cite{a10,a11,a12,a13,a14,a15,a16,a17,a18,a19,a20,a21,a22,a23}.
They are actually
observed in various real systems and are used to model a variety of
processes such as bulk mediated surface diffusion \cite{a17},
exciton and charge transport in polymers under conformational motion
\cite{a18}, transport in micelle systems or heterogeneous rocks
\cite{a19},two-dimensional rotating flow \cite{a20}, and many others\cite{a10}.\\
 \indent
Very few studies on ratchets have focused on the L\'{e}vy flights.
Recently, Dybiec and coworkers \cite{a21} studied the minimal setup
for a L\'{e}vy ratchet and found that due to the nonthermal
character of the L\'{e}vy noise, the net current can be obtained
even in the absence of whatever additional time-dependent forces.
Del-Castillo-Negrete and coworkers \cite{a23} also found the similar
results in constant force-driven L\'{e}vy ratchet for $1<\alpha<2$ ($\alpha$ is L\'{e}vy index).
Rosa and Beims \cite{a24} studied the optimal transport and its relation to superdiffusive transport and
L\'{e}vy walks for Brownian Particles in ratchet potential in the presence of modulated environment and
 external oscillating forces.
 In these studies the L\'{e}vy noise is an intrinsic driving factor
to obtain the net transport. In the classical forced thermal ratchets
\cite{a1,a5}, the driving factor is usually the external ac-driving
force. However, what's the difference between the intrinsic driver
and the external one ? How do L\'{e}vy flights compete with the ac-driving forces ?
In order to answer these questions, in the present paper, we studied the transport of overdamped particles in
an asymmetrically periodic potential in the presence of the L\'{e}vy
flights and ac-driving forces. Our emphasis is on finding the
difference between the two driving factors and how the competition
between them affects the transport.

\indent

\section{Model and Methods}
\indent In this study, we consider the transport of Brownian
particles moving in an asymmetrically periodic potential in the
presence of ac-driving forces and L\'{e}vy-stable noises. The
overdamped dynamics can be described by the following Langevin
equation in the dimensionless form
\begin{equation}\label{}
    \frac{dx}{dt}=-U^{'}(x)+A_{0}\sin(\omega t)+\zeta_{\alpha}(t),
\end{equation}
where $A_{0}$ and $\omega$ are the driving amplitude and frequency,
respectively. The prime stands for differentiation over $x$.
$U(x)$ is an asymmetrically periodic potential (see Fig. 1)
\begin{figure}[htbp]
  \begin{center}\includegraphics[width=8cm,height=4cm]{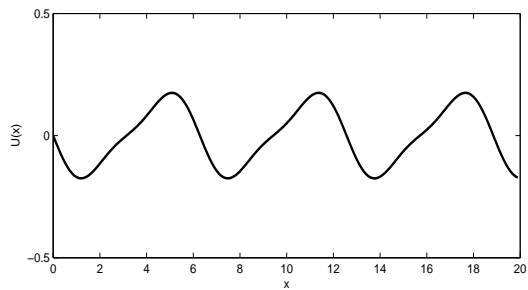}
  \caption{Asymmetrically periodic potential defined by Eq. (2).}\label{1}
\end{center}
\end{figure}
\begin{equation}\label{}
    U(x)=-U_{0}[\sin(x)+\frac{\Delta}{4}\sin(2x)],
\end{equation}
where $U_{0}$ denotes the height of the potential and $\Delta$ is
its asymmetric parameter.

\indent $\zeta_{\alpha}(t)$ is white, symmetric L\'{e}vy-stable
noise with independent increments distributed according to the
stable density with the index $\alpha$. The time integral of the
L\'{e}vy noise over an increment $\Delta t$
\begin{equation}\label{}
    L_{\alpha,\sigma}(\Delta t)=\int_{t}^{t+\Delta
    t}\zeta_{\alpha}(t^{'})dt^{'}
\end{equation}
is an $\alpha$-stable process with stationary independent increments
and its characteristic function (Fourier transform of probability
density function)\cite{a10,a11,a12}
\begin{equation}\label{}
    P_{L}(k,\Delta t)=\exp(-\sigma|k|^{\alpha}\Delta t),
\end{equation}
where $\sigma$ is the intensity of the L\'{e}vy noise and $k$ is wave
number. The parameter $\alpha\in (0,2]$ denotes the stability
index, yielding the asymptotic long tail power law for the
$\zeta$-distribution, which is of the $|\zeta|^{-1-\alpha}$ type.
For the special case $\alpha=2.0$, i.e., for a Gaussian noise, we
are led back to the Brownian case. From Eqs. (3) and (4)
we can obtain the discrete time representation of Eq. (1) for
sufficiently small time step $\Delta t$
\begin{equation}\label{}
    x(t_{n+1})=x(t_{n})-U^{'}(x(t_{n}))\Delta t+A_{0}\sin(\omega t_{n})\Delta
    t+(2D \Delta t)^{\frac{1}{\alpha}}\zeta_{\alpha}(n),
\end{equation}
where $n=0,1,2...$ and $\zeta_{\alpha}(n)$ is a random number
possessing L\'{e}vy stable distribution. In order to compare with
the classical forced thermal ratchets, we use $D=\sigma/2$ to
describe the noise intensity.  In numerical simulations, the
corresponding generator is taken from \cite{a11,a12,b1}
\begin{equation}\label{}
\zeta_{\alpha}(n)=\frac{\sin(\alpha V)}{(\cos
V)^{1/\alpha}}[\frac{\cos([1-\alpha]V)}{W}]^{\frac{1-\alpha}{\alpha}},
\end{equation}
where $V$ is random number uniformly distributed on the interval $(
-\pi/2, \pi/2)$, $W$ is an independent random variable distributed
according to the exponential distribution with unit mean.

\indent In this study, we mainly focus on the transport of the driven
particles. In the Gaussian noise driving ratchets, due to the
existence of the mean value of the noise, the displacement of
particles  also possesses a mean and the transport can be
characterized by the average velocity. However, for the noise with
distribution of a L\'{e}vy-stable law with $0<\alpha<1$, the mean of
the noise and the overall displacement does not exist.
The main feature of this distribution is that the tails cannot be
cut off, or in other words, rare but large events cannot be
neglected. As a consequence, the classical stochastic theory
(average velocity), which is based on the ordinary central limit
theorem, is no longer valid. Recently, Dybiec and coworkers
\cite{a21} proposed a different approach to the L\'{e}vy ratchet
problem based on the group velocity analysis for
$0<\alpha<2 $. Throughout the paper, we will use this method to
describe the transport of the particles.

\indent Median line is a very useful tool for investigation of the
overall motion of the probability density of finding a particle in
the vicinity of $x$ \cite{a11}.  A median line for a stochastic
process $x(t)$ is a function of $q_{0.5}(t)$ given by the
relationship $Pr(x(t)\leq q_{0.5}(t))=0.5$. Therefore, one can use
the derivative of the median to define the group velocity of the
particle packet\cite{a21},
\begin{equation}\label{}
    \upsilon_{g}(t)=\frac{d q_{0.5}(t)}{dt},
\end{equation}
and this definition is valid even for the case of lacking average
current. In the following, we mainly focus on the study of the long
time group velocity,
\begin{equation}\label{}
V_{g}=\lim_{t\rightarrow\infty}\frac{q_{0.5}(t)}{t}.
\end{equation}
   \indent In our simulations, we have considered more than $10^{5}$
   realizations to obtain the accurate median.
   In order to provide the requested accuracy of the system dynamics
   time step was chosen to be smaller than $10^{-3}$. We have
   checked that these are sufficient for the system to obtain
   consistent results.

\section {Numerical results and discussion}

Our emphasis is on finding the median and group velocity with definitions in Eq. (7).
In order to investigate the effects of the
interplay between the ac-driving forces and L\'{e}vy flights we
carried out extensively numerical simulations. For simplicity we set
$U_{0}=1.0$ and $\Delta=1.0$ throughout the work.

\begin{figure}[htbp]
  \begin{center}\includegraphics[width=10cm,height=8cm]{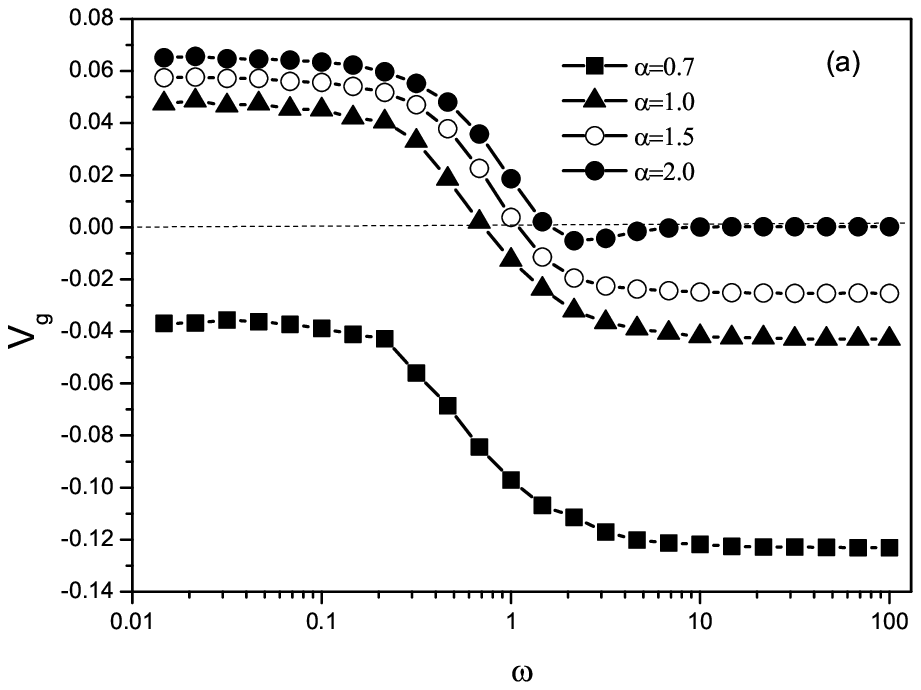}
  \includegraphics[width=10cm,height=8cm]{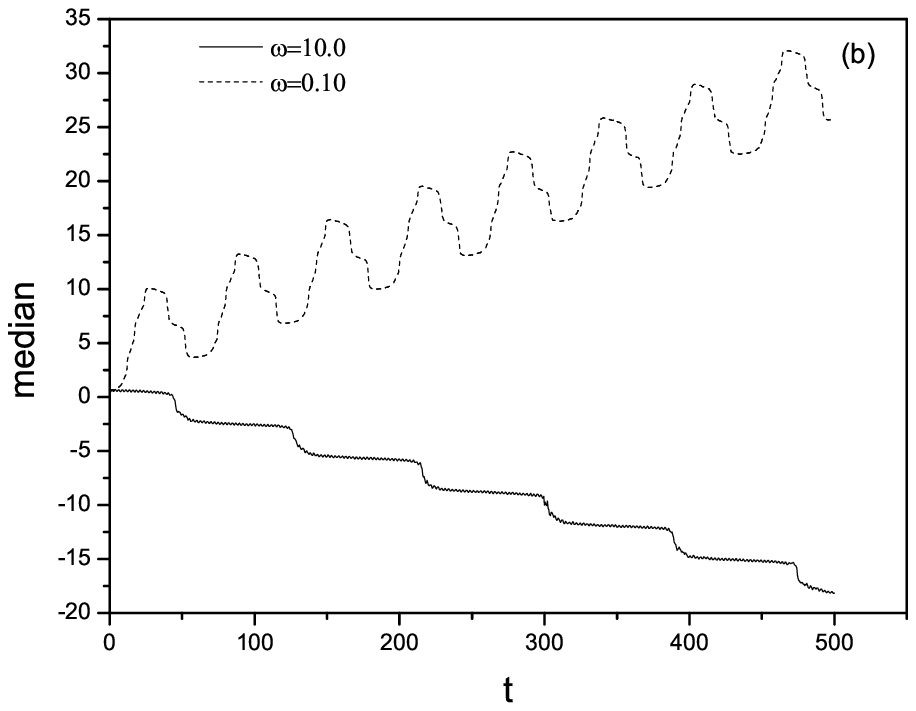}
  \caption{(a)Group velocity $V_{g}$ as a function of the driving frequency $\omega$ for different
  values of L\'{e}vy index $\alpha$ at $D=0.1$ and $A_{0}=1.0$. (b)Time dependence of the location of the median $q_{0.5}(t)$ for
  different values of driving frequency $\omega$ at $D=0.1$ and $A_{0}=1.0$.}\label{1}
\end{center}
\end{figure}
\indent Fig. 2(a) shows the group velocity $V_{g}$ as a function of
the driving frequency $\omega$. For the case of $\alpha=2.0$, the
L\'{e}vy ratchet reduces to the classical forced thermal ratchet
\cite{a1,a5}. In this case, the only resource driving particle
current across the barrier is the ac-driving force. In the adiabatic
limit $\omega\rightarrow 0$, the ac-driving force can be expressed
by two opposite forces $A_{0}$ and $-A_{0}$. The particles get
enough time to cross both side from the minima of the potential. It
is easier for particles to move toward the slanted side than toward
the steeper side, so the group velocity is positive. On increasing
the frequency $\omega$, due to the high frequency, the particles in
one period get more enough time to climb the barrier from the
steeper side than from the slanted side, resulting in negative group
velocity. When the ac-driving forces oscillate very fast, the
particles will experience a time average constant force
$F=\int_{0}^{\frac{2\pi}{\omega}}F(t)dt=0$, so the group velocity
goes to zero. At some intermediate values of $\omega$, the group
velocity crosses zero and subsequently reverse its direction.
However, when $0<\alpha<2$, due to the nonthermal character of
L\'{e}vy noise, L\'{e}vy noise becomes  another source driving the
particle current. As the index $\alpha$ decreases, the positive
group velocity decreases while the negative group velocity
increases. Especially, for $\alpha=0.7$, the L\'{e}vy flights
dominate the transport and the group velocity is even always
negative. It is obvious that the transport driven by the L\'{e}vy
flights is opposite to that driven by the ac-driving forces for not
too small values of the L\'{e}vy noise intensity. In the same
potential, the ac-driving forces will induce a positive group
velocity, while the L\'{e}vy flights will give a negative group
velocity for given index shown in Fig. 2(a). Interestingly, we also
found that group velocity tends to a negative constant, instead of
zero, at fast-driving limit. In this case the effects of ac-driving
forces disappear and the L\'{e}vy flights will dominate the
transport. In Fig. 2(b), we present the time dependence of the
location of the median for a given index $\alpha=1.5$. It is found
that the particles exhibit a motion toward the right direction at
$\omega=0.1$ and opposite direction at $\omega=10.0$.

\begin{figure}[htbp]
  \begin{center}\includegraphics[width=10cm,height=8cm]{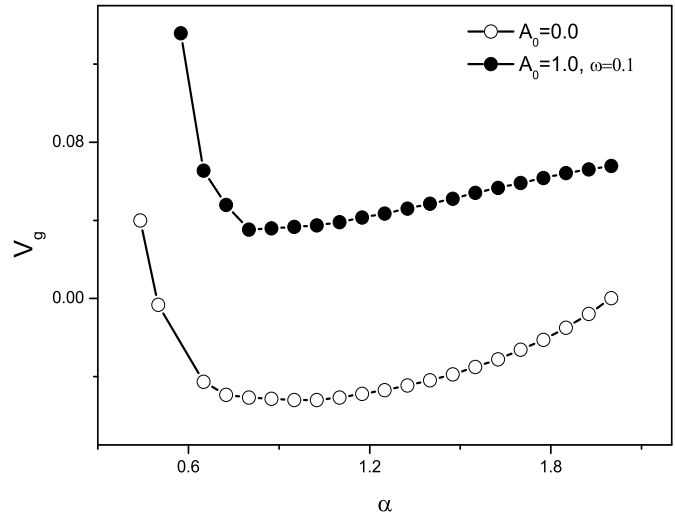}
  \caption{Group velocity $V_{g}$ versus L\'{e}vy index $\alpha$ with and without ac-driving forces at $D=0.1$.}\label{1}
\end{center}
\end{figure}
\indent Figure 3 illustrates the dependence of group velocity
$V_{g}$ on the L\'{e}vy index $\alpha$ with and without ac-driving
forces. From the figure we can see that the rectified transport can
occur even the external driving forces are absent. This is due to
the nonthermal character of the L\'{e}vy noise which can break the
thermaldynamical equilibrium. Interestingly, one can see that the
curves demonstrate nonmonotonic behavior for small values of the
noise intensity. These phenomena can be explained by the interplay
between the potential profile and L\'{e}vy flights. Firstly, the
particles stay in the minima of the potential awaiting large noise
pulses to be catapulted out. For not too small values of L\'{e}vy
index, the L\'{e}vy flights are shorter, and the outliers in the
L\'{e}vy noise are smaller, the distance from minima to maxima
dominates the transport. In this case this distance is shorter from
the steeper side (the left side) than that from the slanted side
(the right side). Consequently, most of the particles are thrown out
from the steeper side, resulting in negative group velocity.
However, for very small values of L\'{e}vy index, the L\'{e}vy
flights are longer, and the outliers in the L\'{e}vy noise are
larger. In this case, the slope of the potential dominates the
transport. So it is easier for the particles moving toward the
slanted side than toward the steeper side and the group velocity is
positive. Therefore, there exists an intermediate value of $\alpha$
at which the group velocity takes its extra value and the group
velocity as a function of the L\'{e}vy index is nonmonotonic.
However, for large values of the noise intensity, the curves
demonstrate monotonic behavior. This case was reported in Ref.[21].
When a large ac-driving force ($A_{0}=1.0$) is added, the ac-driving
force dominates the transport and the group velocity is positive.
However, the shape of the curve is similar to that without the
ac-driving forces.

\begin{figure}[htbp]
  \begin{center}\includegraphics[width=10cm,height=8cm]{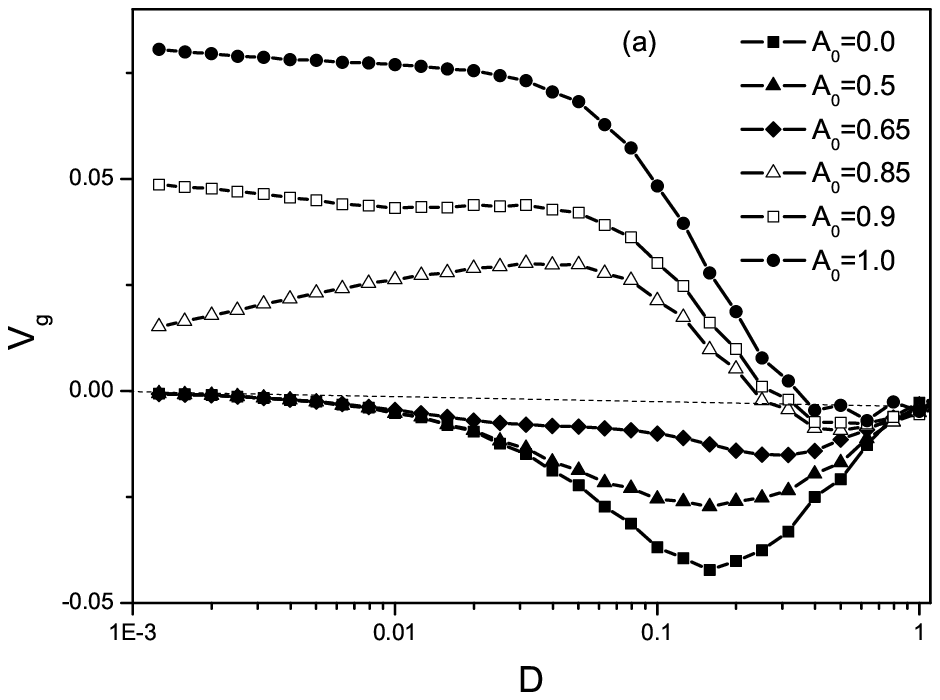}
  \includegraphics[width=10cm,height=8cm]{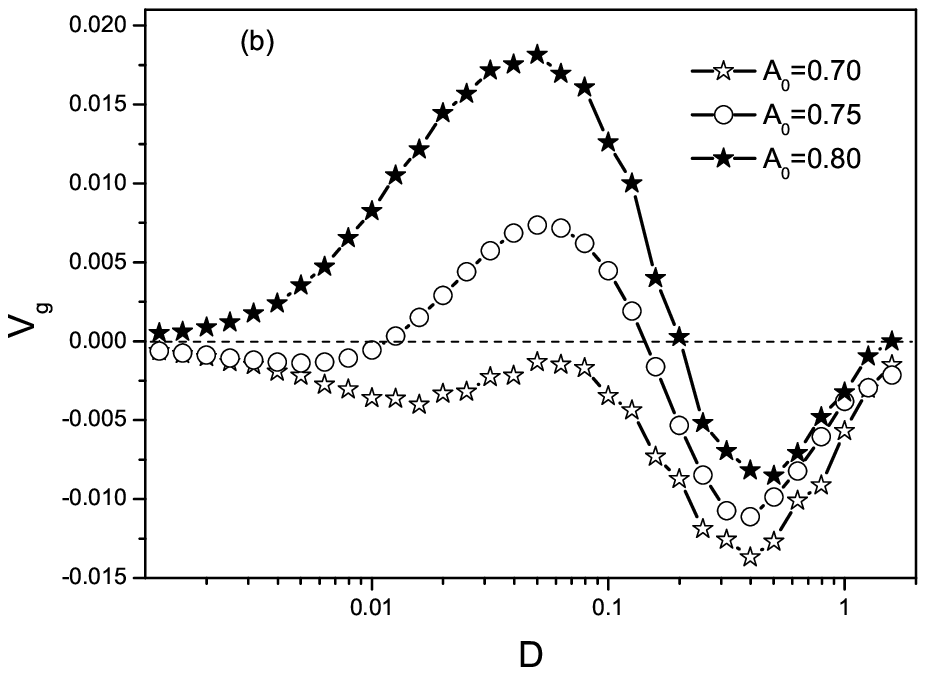}
  \caption{Group velocity $V_{g}$ as a function of the noise intensity $D$ at $\alpha=1.5$ and $\omega=0.1$.
  (a) $A_{0}$=$0.0$, $0.5$, $0.65$, $0.85$, $0.9$, and $1.0$. (b) $A_{0}$=$0.70$, $0.75$, $0.80$.}\label{1}
\end{center}
\end{figure}
\indent In Fig. 4, the group velocity $V_{g}$ is plotted for
different values of amplitude $A_{0}$ as a function of the noise
intensity $D$. When the driving amplitude $A_{0}$ is small, the
L\'{e}vy flights will dominate the transport and the group velocity
is negative. On increasing the amplitude $A_{0}$, the ac-driving
forces gradually dominate the transport and the group velocity
crosses zero and becomes positive, namely, transport reversal
occurs. When $D\rightarrow 0$, the group velocity tends to zero for
small driving fores ($A_{0}=0.0$, $0.5$, $0.65$, and $0.75$) and
goes to a finite value for large driving forces ($A_{0}=0.85$,
$0.9$, and $1.0$). This is due to the fact that in the determined
ratchets the net current occurs for large driving forces and disappears
for small driving forces. When $D\rightarrow
\infty$, the effects of the potential and the ac-driving forces
disappear, so the group velocity tends to zero. For
suitable amplitude $A_{0}$ the group velocity can change its
direction on increasing the noise intensity. Remarkably, multiple
transport reversals even occur at $A_{0}=0.75$ (see Fig. 4(b)) and the group
velocity reverses its direction twice. The intensive competition
between the ac-driving forces and the L\'{e}vy flights leads to this
peculiar phenomenon.

\begin{figure}[htbp]
  \begin{center}\includegraphics[width=10cm,height=8cm]{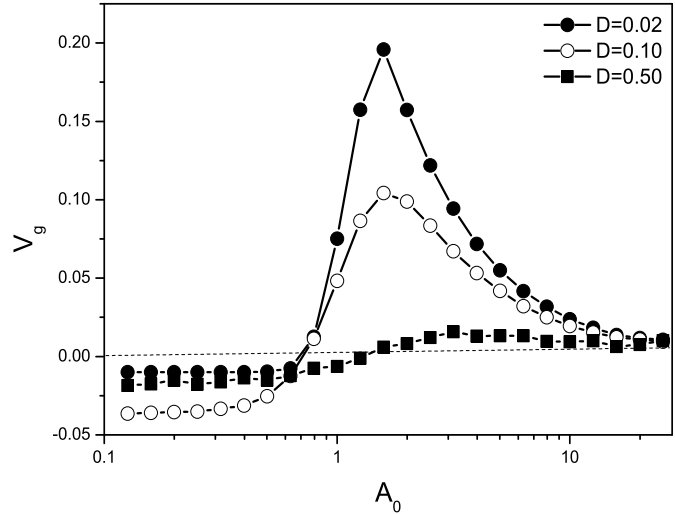}
  \caption{Dependence of group velocity $V_{g}$ on the driving amplitude $A_{0}$ for different values of noise intensity at $\alpha=1.5$ and $\omega=0.1$.}\label{1}
\end{center}
\end{figure}

\indent We next investigate the role of the amplitude $A_{0}$ on the
transport. The results are shown in Fig. 5.  It is found that the
group velocity will tend to a negative constant  for very small
driving forces. As we know that in a classical forced thermal ratchet \cite{a1,a5} the velocity will tend to zero for very small driving forces.
The nonthermal character of the L\'{e}vy noise induces this
different behavior. For very large values of the amplitude, the
influence of the potential and the L\'{e}vy flights will become
negligible and the group velocity goes to zero. At some intermediate
values of amplitude $A_{0}$, the group velocity crosses zero and
reverses it direction.

\section{Concluding Remarks}
\indent In this paper, we studied the transport of overdamped
particles moving in an asymmetrically periodic potential in the
presence of ac-driving forces and L\'{e}vy flights. Because the mean
of the L\'{e}vy noise and the mean of the overall displacement do
not exist, the definition of the average velocity may be invalid.
Therefore, throughout the study we use the group velocity proposed
by Dybiec and coworkers \cite{a22} to measure the transport of
overdamped particles. From the Langevin numerical simulations, we
found that ac-driving forces and L\'{e}vy flights are the two
different drivers in nature that can break thermodynamical
equilibrium. The competition between the two driving factors induces
some peculiar phenomena. Due to the L\'{e}vy flights, the group
velocity tends to a negative constant, instead of zero, for
fast-driving limit and it may be always negative for small value of
the L\'{e}vy index. We also found that the relation between the
group velocity and L\'{e}vy index is nonmonotonic for small values
of the noise intensity. There exists an intermediate value of
L\'{e}vy index at which the group velocity takes its extreme value.
Remarkably, multiple transport reversals occur when the noise
intensity changes. This is caused by the competition between the
ac-driving forces and L\'{e}vy flights.

 \indent Though the model presented does not pretend to be a realistic model for
 a real system, beyond its intrinsic theoretical interest, the results we have presented
 have potential applications in many processes such as diffusive transport in plasmas,
 particles separation with non-Gaussian diffusion, and ratchet transport
 in biology systems that are intrinsically out of equilibrium.

\indent  This work was supported in part by National Natural Science
Foundation of China with Grant No. 30600122 and GuangDong Provincial
Natural Science Foundation with Grant No. 06025073. Y. F. He also
acknowledges the Research Foundation of Education Bureau of Hebei
Province, China (Grant No. 2009108)

\end{document}